# MUSIC: A Hybrid Computing Environment for Burrows-Wheeler Alignment for Massive Amount of Short Read Sequence Data

Saurabh Gupta[1], Sanjoy Chaudhury[1] and Binay Panda[1,2]
[1]Bio-IT Centre, Ganit Labs, Institute of Bioinformatics and Applied Biotechnology
[2]Strand Life Sciences
Bangalore, India
binay@ganitlabs.in

*Abstract*—High-throughput DNA sequencers are becoming indispensable in our understanding of diseases at molecular level, in marker-assisted selection in agriculture and in microbial genetics research. These sequencing instruments produce enormous amount of data (often terabytes of raw data in a month) that requires efficient analysis, management and interpretation. The commonly used sequencing instrument today produces billions of short reads (upto 150 bases) from each run. The first step in the data analysis step is alignment of these short reads to the reference genome of choice. There are different open source algorithms available for sequence alignment to the reference genome. These tools normally have a high computational overhead, both in terms of number of processors and memory. Here, we propose a hybrid-computing environment called MUSIC (Mapping USIng hybrid Computing) for one of the most popular open source sequence alignment algorithm, BWA, using accelerators that show significant improvement in speed over the serial code.

I. INTRODUCTION

Next generation sequencing instruments generate massive amounts of data, to the order of 600 gigabytes of raw base-call data, in a single run that can take as long as 12 days. Although there are multiple sequencing instruments available in the market today, the HiSeq instrument from Illumina is the most commonly used instrument. The sequencing instrument produces raw base calls from the image data for all the four possible nucleotides, A, T, G and C. Aligning these raw base call data, arranged in a string of upto 150 letters (a letter representing either A, T, G or C) to a reference genome is the first step towards data analysis. Although there are several open source tools available for sequence alignment [1-15] for short-read sequence data, Burrows-Wheelers Alignment (BWA) [1] remains the most popular choice among researchers. BWA is based on the concept of string matching originally developed by Burrows and Wheeler [15]. Most alignment algorithms are based either on hash-table or suffix-trees approach [16]. The speed, memory footprint and computational hardware requirement are major drawbacks of the majority of currently used sequence alignment tools, making them unsuitable for individual biology labs that normally are not equipped with sophisticated hardware resources and computational know-how. Additionally, the hardware requirement for different applications and genomes are different. For example, a lab that is studying viral genomes may not need a sophisticated multi-node high performance computing (HPC) solution where a few quad-core instruments can do the requisite job. In contrast, a lab studying human diseases, like cancer, and routinely sequences tens and hundreds of genomes, exomes and transcriptomes requires rather a large HPC solution.

Lately, there has been a major interest in using Graphics Processing Units (GPUs) for computationally intense scientific work [17]. GPUs are designed based on the premise that latency (speed) can be compromised for gain in throughput. Unlike CPUs, GPUs have large processing elements to cache space ratio. This gives GPUs hardware multithreading and single instruction multiple data (SIMD) execution ability. Real time computer graphics require a high throughput hardware platform to implement, as the number of parallel calculations to be done is high due to huge number of pixels [17]. Therefore GPUs have massively parallel high throughput architecture. Since sequence alignment algorithms consist of a single set of instructions, mapping short reads to a reference genome, they are ideal candidates for implementation on high throughput architectures like GPUs [16]. Additionally, short-read mapping to large reference genomes, such as those from some plants, can take considerable CPU hardware resources and long time to align where GPU-based alignment algorithms may come useful.

Sequence alignment algorithms were implemented using GPUs in the past [18, 19] including for BWA [20, 21]. We tried a similar approach by using a single NVIDIA Tesla C2070 and single and multiple NVIDIA K20 GPU accelerators and obtained significant improvement in speed over the CPU code. We tested our code both for high coverage simulated and multiple real biological data.

## II. GPU IMPLEMENTATION

The BWA CPU code is implemented in 3 stages:
- Indexing (to create the BWT, Suffix Array or SA, C and Occ arrays and save them as intermediate files).
- Alignment (to map the reads against the reference string using the data structures created by the indexing step, to get the SA intervals).
- The final step where the alignments are written in SAM format.

Here, we focus only on the "Alignment" step as it is the most time consuming step in the BWA algorithm. Hence, our implementation uses the BWA's serial code for performing indexing and generating alignments in SAM format, whereas the alignment step is carried out using the GPU accelerators.

TABLE I. SPECIFICATIONS OF GPU ACCELERATOR USED.

| A. Instrument 2: NVIDIA Tesla C2070 ||
|---|---|
| Property | Details |
| Number of CUDA cores | 448 |
| Frequency of each core | 1.15GHz |
| Total dedicated memory | 6GB GDDR5 |
| Memory speed | 1.5GHz |
| Memory interface | 384-bit |
| Memory bandwidth | 144GB/s |

| B. Instrument 2: NVIDIA Tesla K20 ||
|---|---|
| Property | Details |
| Number of CUDA Cores | 2496 |
| Frequency of each core | 1.15GHz |
| Total dedicated memory | 5GB GDDR5 |
| Memory speed | 1.5GHz |
| Memory interface | 384-bit |
| Memory bandwidth | 208GB/s |

We used NVIDIA's Compute Unified Device Architecture (CUDA) software development environment to execute MUSIC by using a single Tesla C2070 card. The CPU used to run the serial code was Intel core i7 CPU 960 3.20GHz instrument that used a Linux Operating System. We attempted parallelizing the alignment step in such a way that copies the reference sequence and input reads from CPU to GPU for the CPU-GPU load balance and then launches the multiple concurrent kernels for handling the large number of sequence reads harnessing the massive numbers of GPU processors. We also tried using texture memory for the fast data access. We tested MUSIC both on simulated (human chr22, 6.9m reads) and multiple real biological data (three different and independent human whole exomes with 100m reads). Detailed specifications of the instrument on which the CUDA code was executed is given in Table 1.

The workflow for MUSIC is presented in Fig. 1. Different steps involved are:

- The input reads are read first from file to memory by the CPU.
- The CPU index then lookups to find positions in the genome.
- The GPU aligns the reads to the available positions & finds the best match
- The CPU processes the output from the GPU.
- And finally the CPU writes the output to a file.

We divided the input sequence reads into short fragments of 32 *mers* (default seed length) and performed alignment by multiple consecutive kernel runs. The alignment data for each of the sequence reads, including BWT SA coordinates and the number of differences, was stored temporarily in the GPU memory. Once the kernel finished the job, the data were copied from the GPU back to the host CPU and subsequently written onto the disk storage in a binary file format.

## III. RESULTS AND DISCUSSION

After the GPU implementation, we compared the runtimes of GPU codes for simulated and real multiple biological data. The CPU code was run using an Intel core i7 CPU 960 3.20GHz under the Linux Operating System and the GPU on NVIDIA Tesla C2070. First, we wanted to see if MUSIC performs better than the CPU code alone, as we had originally expected. In order to do this, we used the high coverage human exome data (total of 100m reads) and obtained significant time improvement with when we compared the performance of MUSIC with the CPU code alone (Fig. 2). Next, we wanted to compare the performance of MUSIC with that with other GPU-enabled alignment algorithms published earlier. As shown in Fig. 3A, MUSIC performs at par with some other published algorithm (barracuda) and significantly better than others (cushaw) previously implemented with GPU accelerators. We further extended this observation with multiple real biological data and obtained the expected results (Fig. 3B).

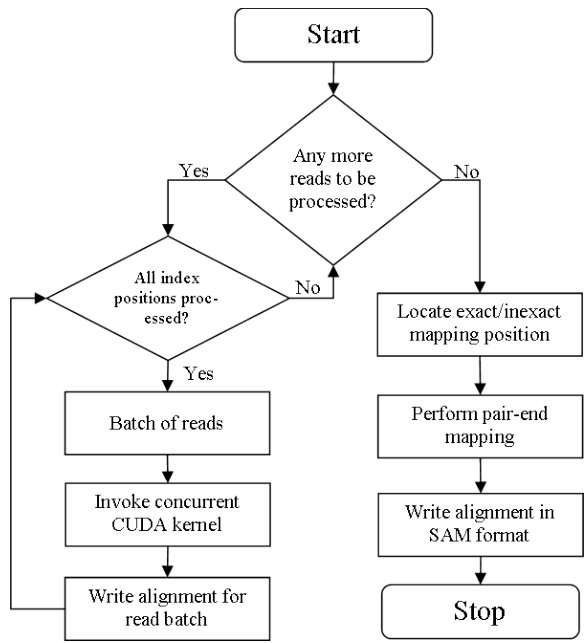

Fig. 1. MUSIC workflow.

Finally, we wanted to test a different, a more recent, NVIDIA GPU accelerator to see if we could even further improve the performance of MUSIC. As the previous GPU-enabled alignment algorithms used the older set of GPU accelerators (C2050) and the memory bandwidth and the number of cores per GPU in both the C2050 and C2070 accelerators are the same, we wanted to test whether we could further improve our results using the kepler K20 accelerators that has more number of cores and better memory bandwidth. As shown in Fig. 4, we got significant improvement of speed when we used either one (blue bars) or two (red bars) kepler K20 GPU accelerators using multiple real biological data.

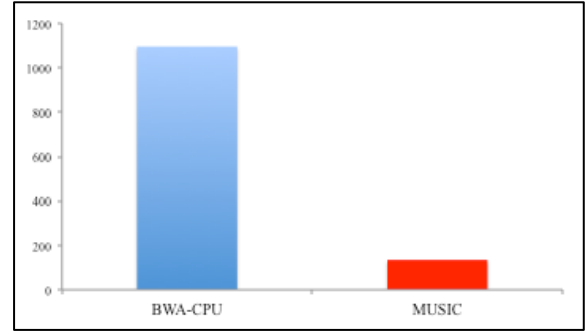

Fig. 2. Comparison of CPU (BWA-CPU) vs GPU (MUSIC), Y-axis: total time (alignment + sampe) in min.

This improvement in alignment time in the GPU-enabled tools has come about through better implementation of the memory hierarchy and crucial use of the texture variable to maximize the thread efficiency. The *ab intio* approach has resulted in performance improvement, by keeping our implementation simple we have not sacrificed clarity. By implementing one of the most popular alignment algorithm using GPU opens the possibility for a large number of smaller biology-driven labs, which have less or no access to HPC solutions, use high-throughput sequencing data.

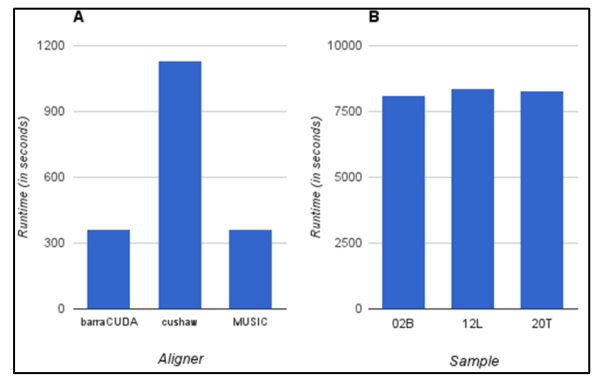

Fig. 3. Performance comparison between MUSIC and other GPU-enabled tools using simulated (A) data and that of MUSIC using multiple real biological data (B), Y-axis: total time (alignment + sampe).

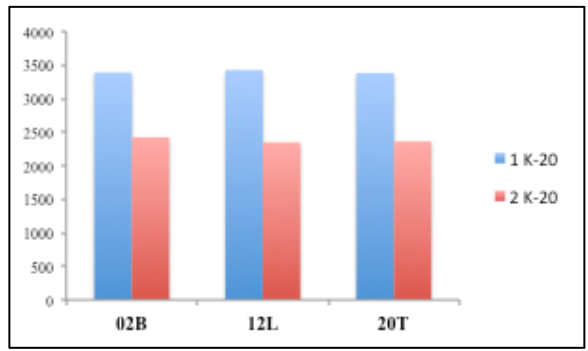

Fig. 4. Performance comparison of MUSIC using either a single (blue bars) or two (red bars) Tesla K20 GPU accelerators, Y-axis: time in sec.

Future improvement in MUSIC will require refinement of code to making multi-GPU implementation work better in terms of alignment time, using recent and more powerful GPU accelerators, and making the program easy to use for biologists.

ACKNOWLEDGMENT

Research is funded by the Department of Electronics and Information Technology, Government of India (Ref No:18(4)/2010-E-Infra., 31-03-2010) and Department of IT, BT and ST, Government of Karnataka, India (Ref No:3451-00-090-2-22) under the "Bio-IT Project". The GPU accelerators C2070 used in this paper were donated by NVIDIA under a 'Professor Partnership Program' to BP and the server time with K20 accelerators was donated to BP from NVIDIA.